\documentclass[journal]{IEEEtran}
\usepackage{multicol}
\usepackage{subfigure}
\usepackage{amsmath}
\usepackage{amssymb}
\usepackage{amsfonts}
\usepackage{graphicx}
\usepackage{url}
\usepackage{ccaption}
\usepackage{booktabs}

\usepackage{cite}

\ifCLASSINFOpdf
\else
\fi
\begin{document}
%
\title{A Novel Carrier Waveform Inter-Displacement Modulation Method in Underwater Communication Channel}
%
%
%

\author{Hai-Peng~Ren,{}
        and~Yang~Zhao{}
\thanks{}
\thanks{}
\thanks{}}

%
%

\markboth{}%
{Shell \MakeLowercase{\textit{et al.}}: Bare Demo of IEEEtran.cls for Journals}
%



\maketitle

\begin{abstract}
As the main way of underwater wireless communication, underwater
acoustic communication is one of the focuses of ocean research.
Compared with the free space wireless communication channel, the
underwater acoustic channel suffers from more severe multipath
effect, the less available bandwidth and the even complex noise. The
underwater acoustic channel is one of the most complicated wireless
communication channels. To achieve a reliable underwater acoustic
communication, Phase Shift Keying (PSK) modulation and Passive Time
Reversal Mirror (PTRM) equalization are considered to be a suitable
scheme. However, due to the serious distortion of the received
signal caused by the channel, this scheme suffers from a high Bit
Error Rate (BER) under the condition of the low Signal to Noise
Ratio (SNR). To solve this problem, we proposes a Carrier Waveform
Inter-Displacement (CWID) modulation method based on the Linear
Frequency Modulation (LFM) PSK and PTRM scheme. The new
communication scheme reduces BER by increasing the difference from
the carrier waveform for different symbols. Simulation results show
the effectiveness and superiority of the proposed method.
\end{abstract}


%
\IEEEpeerreviewmaketitle

\section{Introduction}
%
%
%
%
\IEEEPARstart{T}{he} ocean covers about 360 million square
kilometers occupying $71\%$ of the earth's surface. The ocean
resources are important for human being. With the development of
ocean commercial exploitation and ocean defense, a growing interest
in underwater communication has been witnessed. Underwater cable
communication suffers from high cost and inconvenience. Therefore,
underwater wireless communication are considered as a better way of
underwater communication. Radio waves have a severe attenuation when
being transmitted in the water. Acoustic waves are the optimal
carrier wave in underwater wireless communication
\cite{Stojanovic2008}. As the main way of underwater wireless
communication, underwater acoustic communication becomes more and
more important in the field of ocean research
\cite{Shahabudeen2008}.

Underwater acoustic channel is recognized as one of the most complex
communication channels. Compared with the free space wireless
communication channel, the underwater acoustic channel has a more
severe multipath effect, a less available bandwidth, a more serious
Doppler spreading and shifting, even a more complex
noise\cite{Stojanovic2008,Du2011,Stojanovic2009}. The suitable
modulation and equalization are necessary to implement a high speed
and low BER underwater communication.

Multi-carrier modulation \cite{Li2008} and single-carrier modulation
are the two main modulation methods today. A higher speed can be
achieved by using multi-carrier modulation. However, the modulation
method is sensitive with the frequency offset of the carrier-waves,
which leads to a high BER in a poor environment \cite{He2010}.
Single-carrier modulation involves incoherent modulation and
coherent modulation. Incoherent modulation was widely used in the
early stage of underwater acoustic communication, for example
Frequency Shift Keying (FSK) modulation. The advantages of
incoherent modulation are simplicity and reliability. However,
incoherent modulation are limited by a low bandwidth efficiency.
With the increasing request of improving bandwidth efficiency,
coherent modulation was used \cite{Stojanovic2008}. PSK modulation
is the most widely used coherent modulation, which uses different
initial phases to represent different symbols. Binary Phase Shift
Keying (BPSK) modulation has a low BER due to the large difference
of carrier-waves for different symbols. However, its bit
transmission rate is low. M-ary Phase Shift Keying (MPSK) modulation
can get a higher transmission bit rate, but with a higher BER. For
example, for Quadrature Phase Shift Keying (QPSK) modulation, the
speed is doubled compared with that of BPSK. However, the
differences between carrier-waves of different symbols are reduced,
it causes an increasing BER.

Due to the complexity of the underwater acoustic channel, the
signals passing through it suffer from a more severe distortion
compared to its counterpart in a free space wireless communication
channel. Equalization is a one of the techniques to resist the
distortion. Adaptive equalizations are the widely used equalizations
in underwater acoustic communication
\cite{Stojanovic1993,Rouseff2001,Yin2008,Vadde2011}. The underwater
acoustic communication channel exhibits a long delay spread because
of a large amount of multipath arrivals resulting from surface and
bottom interactions. The long delay spread and rapid time variation
of the channel causes the computation of adaptive decision feedback
equalization too complex for real-time implementations
\cite{Shahabudeen2008}. Therefore, several advanced algorithms based
on the structure of decision feedback equalizer were proposed, for
example, a decision feedback equalizer coefficient placement
algorithm for sparse reverberant channel \cite{Lopez2001}, an
adaptive decision feedback equalizer based on composed VFF-RLS and
VSS-NLMS algorithm \cite{Zhao2010}, a receiver using adaptive
decision feedback equalizer and Bit Interleaved Coded Modulation
with Iterative Decoding (BICM-ID) \cite{Shah2011}. Compared with the
original adaptive decision feedback equalization, these modified
algorithms have lower computation consumption and better
performance. However, they all suffer from error propagation due to
the feedback of erroneous decisions in the loop. PTRM equalization
can match the underwater acoustic channel automatically without any
transcendental knowledge. PTRM equalization also reduces the
computational computation.

The communication schemes using PSK modulation and PTRM equalization
are recognized as a suitable underwater wireless communication
scheme. An experiment was carried out in water tank using BPSK
modulation \cite{Lu2005}, 7 chips Barker code, and PTRM
equalization. The results showed that PTRM can decrease BER and
increase communication distance. The BPSK and QPSK modulation
together with PTRM were tested in \cite{Edelmann2005}, the results
showed that the QPSK obtained a higher speed but a higher BER
compared with BPSK. A SISO Irregular Repeat Accumulate (IRA) channel
decoder was proposed to use together with Differential Binary Phase
Shift Keying (DBPSK) modulation and PTRM equalization to decrease
the BER in \cite{Keeser2009}. A spatial diversity referred to as
adaptive spatial combining was proposed to use together with QPSK
modulation and PTRM equalization to increase the transmission speed,
as a result, 2Kbps transmission rate was obtained in
\cite{Zhang2011}. A multiuser underwater acoustic communication
using 16 quadrature amplitude modulation (16QAM) and PTRM
equalization was implemented in \cite{Song2010} and each user could
get a transmission speed of 2Kbps. Using M-ray modulation to replace
BPSK modulation could increase the transmission rate, but the BER
increases simultaneously. To solve this problem, a new modulation
method called CWID modulation is proposed in this paper. The main
idea of the method is to increase the difference between different
symbols' carrier waveform. The proposed method possesses high
transmission rate and low BER.

This paper is organized as follows: Section 2 introduces the
underwater acoustic channel model; Section 3 gives the principle of
underwater acoustic communication schemes using QPSK modulation and
PTRM equalization; Section 4 proposes underwater acoustic
communication schemes using CWID modulation and PTRM equalization;
Section 5 shows the results of the simulations; Section 6 concludes
the work.
\section{Underwater acoustic channel model}
\subsection{Physical Features of Underwater Acoustic Channels}
Underwater acoustic channel has unique channel features such as
extended multipath, severe Doppler spreading and shifting, complex
noise and less available bandwidth.

Extended multipath, which caused by sound reflection at the surface,
bottom, any objects, and sound refraction in the water, is
recognized as the main challenge in underwater acoustic
communication. The multipath effect in underwater acoustic channel
is much more serious than that in free space wireless channel due to
the transmission speed of the sound is about 1500m/s in the water. A
signal that travels along different paths and arrives many times at
the receiving location will cause serious interference that make the
difficulty to demodulate the information in it
\cite{Du2011,Stojanovic2009}.

The Doppler spreading and shifting in underwater acoustic
communication, which caused by relative motion between the
transmitter and receiver as well as drifting with waves, currents
and tides, has to be taken into account compared with that in free
space wireless communication \cite{Stojanovic2009}. The Doppler
spreading and shifting in underwater acoustic communication is more
serious than that in free space wireless communication due to the
speed of sound (1500m/s) is much slower than the speed of
electro-magnetic wave ($3\times{10^8}$m/s).

The noise consists of ambient noise and site-specific noise in
underwater acoustic channel. The ambient noise comes from sources
such as turbulence, breaking waves, rain and distant shipping.
Site-specific noise exists only in certain places, for example, ice
cracking noise occurs only in polar regions. Furthermore, the noise
may change with the environment, which increases the complexity of
the noise in water \cite{Stojanovic2009}.

Attenuation of the signal transmitted in underwater acoustic channel
increases rapidly with the increase of signal frequency, which leads
to a limited available bandwidth \cite{Stojanovic2009}. Therefore,
the methods with high bandwidth efficiency are expected in
underwater acoustic communication.
\subsection{The Channel Model}
The geometric model of the underwater acoustic channel is showed in
Fig. 1. The mathematical model of the underwater acoustic channel
including multipath, attenuation, Doppler spreading and shifting,
and noise are given as follows \cite{Chitre2007}.
\begin{figure}[!t]
\centering
\includegraphics[width=2.5in]{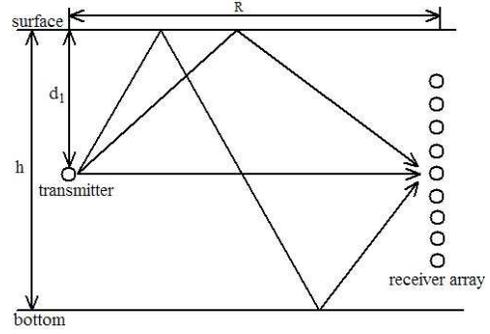}
\caption{Geometric Model of the Underwater Acoustic Channel.}
\label{fig_sim}
\end{figure}

Assume that there are three eigenrays in the communication.
Calculate the distance of each eigenray as follows. Assume that
${D_1}$ is the distance of the direct eigenray, ${D_2}$ is the
distance of the eigenray reflected by surface only, and ${D_3}$ is
the distance of the eigenray reflected by both surface and bottom.
We have
\begin{equation}
\label{eqn_example} {D_1} = \sqrt {{R^2} + {{({d_1} - {d_2})}^2}}
\end{equation}
\begin{equation}
\label{eqn_example} {D_2} = \sqrt {{R^2} + {{({d_1} + {d_2})}^2}}
\end{equation}
\begin{equation}
\label{eqn_example} {D_3} = \sqrt {{R^2} + {{(2h + {d_1} -
{d_2})}^2}}
\end{equation}
where $R$ is the transmission distance, ${d_1}$ is the depth of the
transmitter, ${d_2}$ is the depth of the receiver, $h$ is the height
of the water column.

Transmission loss consists of the loss caused by pressure amplitude
due to spherical spreading, denoted as ${L_{ss}}$, the loss caused
by continuously transformed into heat, denoted as ${L_A}$, the loss
caused by the interaction with surface, denoted as ${L_s}$, and the
loss caused by the interaction with bottom, denoted as ${L_b}$.
\begin{equation}
\label{eqn_example} {L_{ss}} = \frac{1}{D}
\end{equation}
\begin{equation}
\label{eqn_example} {L_A} = \exp \left[ { -
0.998D(\frac{{SA{f_T}{f^2}}}{{f_T^2 + {f^2}}} +
\frac{{B{f^2}}}{{{f_T}}})} \right]
\end{equation}
where $A$=0.00000234, $B$=0.00000338, $S$ is the salinity of the
seawater, $T$ is the temperature of the seawater, $f$ is the
frequency of the carrier-waves, ${f_T}$ is a relaxation frequency
\cite{Brekhovskikh2002}.
\begin{equation}
\label{eqn_example} {f_T} = 2.19 \times {10^{6 - \frac{{1520}}{{T +
273}}}}
\end{equation}
\begin{equation}
\label{eqn_example} {L_s} =  - 1
\end{equation}
\begin{equation}
\label{eqn_example} {L_b} = \left| {\frac{{m\cos \theta  - \sqrt
{{n^2} -{{\sin}^2}\theta } }}{{m\cos \theta  + \sqrt {{n^2}
-{{\sin}^2}\theta }}}} \right|
\end{equation}
where
\begin{equation}
\label{eqn_example} m = \frac{{{\rho _1}}}{\rho },n =
\frac{c}{{{c_1}}}
\end{equation}
\begin{equation}
\label{eqn_example} \theta  = {\tan ^{ - 1}}\left( {\frac{R}{{2bh +
{d_1} - {{( - 1)}^{s- b}}{d_2}}}} \right)
\end{equation}
$\rho$ and $c$ is the density and sound speed in seawater,
respectively. ${\rho_1}$ and ${c_1}$ is the density and sound speed
in the seabed, respectively. $s$ is the reflection times with
surface, $b$ is the reflection times with bottom. Let ${\tau_1}$ be
the arrival time of eigenray ${D_1}$, ${\tau_2}$ be the arrival time
of eigenray ${D_2}$, ${\tau_3}$ be the arrival time of eigenray
${D_3}$. We have
\begin{equation}
\label{eqn_example} {\tau _1} = \frac{{{D_1}}}{c},{\tau _2} =
\frac{{{D_2}}}{c},{\tau _3} = \frac{{{D_3}}}{c}
\end{equation}

As the time-varying features of the underwater acoustic channel
should be taken into account, some statistical variables are
introduced in the model. $A(t)$, representing the fading of
individual eigenpath, are modeled as independent Rayleigh processes
with unit mean and an exponential autocorrelation specified by the
Doppler spreading. $J(t)$, representing the time jitter, are modeled
as Gaussian processes with zero mean, variance ${\sigma^2}$ and an
exponential autocorrelation specified by a transducer position
coherence time. In details, the $A(t)$ and $J(t)$ are fixed in the
coherence time and altered out of the coherence time, namely, the
$A(t)$ and $J(t)$ are same in the same data stream and different in
the different data streams.

The noise in the model is considered as white noise, which denoted
as $n(t)$.

The mathematical model of the underwater acoustic channel can be
concluded as
\begin{equation}
\label{eqn_example} y(t) = \sum\limits_{i = 1}^n
{{A_i}(t){L_{ssi}}{L_{Ai}}L_{si}^sL_{bi}^bx(t - {\tau _i} -
{J_i}(t))}  + n(t)
\end{equation}
where $x(t)$ is the transmitted signal, $y(t)$ is the received
signal.
\section{Underwater acoustic communication based on QPSK and PTRM}
The schematic diagram of the underwater acoustic communication
system based on QPSK modulation and PTRM equalization is showed in
Fig. 2. The system consists of QPSK modulation, PTRM equalization
and matched filter demodulation.
\begin{figure}[!t]
\centering
\includegraphics[width=3.2in]{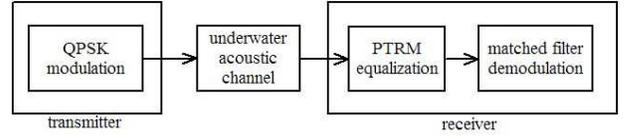}
\caption{Underwater Acoustic Communication Schematic Diagram Based
on QPSK and PTRM.} \label{fig_sim}
\end{figure}
\subsection{QPSK Modulation}
The main idea of the QPSK modulation is using the different initial
phases to represent the different symbols \cite{Goldsmith2005}. The
relationship between initial phases and corresponding symbols is
showed in Table I. Gray coding has been considered for the QPSK
symbols in Talbe I. The carrier-waves of the modulation can be
either cosine signals or LFM signals.
\begin{table}[!t]
\renewcommand{\arraystretch}{1.3}
\caption{QPSK Modulation} \label{table_example} \centering
\begin{tabular}{c||c}
\hline
\bfseries Initial phases & \bfseries Symbols\\
\hline\hline
0                                                                                            &   00   \\
${\pi  \mathord{\left/ {\vphantom {\pi  2}} \right.\kern-\nulldelimiterspace} 2}$            &   01   \\
$\pi$                                                                                        &   11   \\
${{3\pi } \mathord{\left/ {\vphantom {{3\pi } 2}} \right.\kern-\nulldelimiterspace} 2}$      &   10   \\
\hline
\end{tabular}
\end{table}
\subsubsection{QPSK Modulation Based on Cosine Carrier-waves}
The four carrier-waves of the QPSK modulation based on cosine
carrier-waves are showed in Fig. 3. The frequency of the cosine
carrier-waves is 11.5kHz. In Fig. 3, we can find that the only
difference among the four carrier-waves is the initial phases,
showed in Table I.
\begin{figure}[!t]
\centering
\includegraphics[width=3.5in]{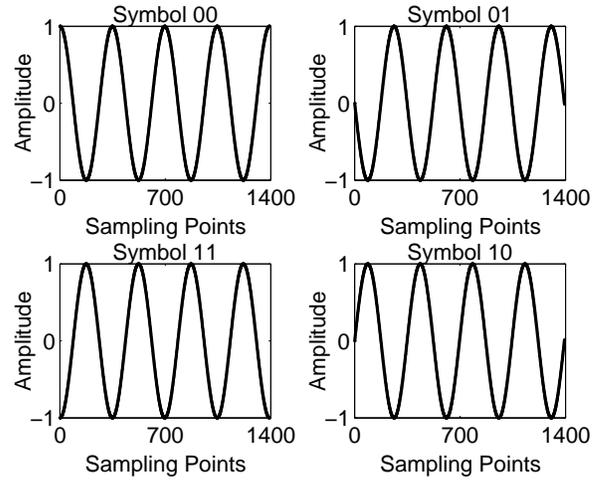}
\caption{QPSK Modulation Based on Cosine Carrier-waves.}
\label{fig_sim}
\end{figure}
\subsubsection{QPSK Modulation Based on LFM Carrier-waves}
The four carrier-waves of the QPSK modulation based on LFM are
showed in Fig. 4. The frequency of the LFM carrier-waves is 5-18kHz.
We can find that the difference between the LFM carrier-waves of
different symbols is greater than that of cosine carrier-waves,
which decreases BER of the communication system compared with the
system using cosine wave.
\begin{figure}[!t]
\centering
\includegraphics[width=3.5in]{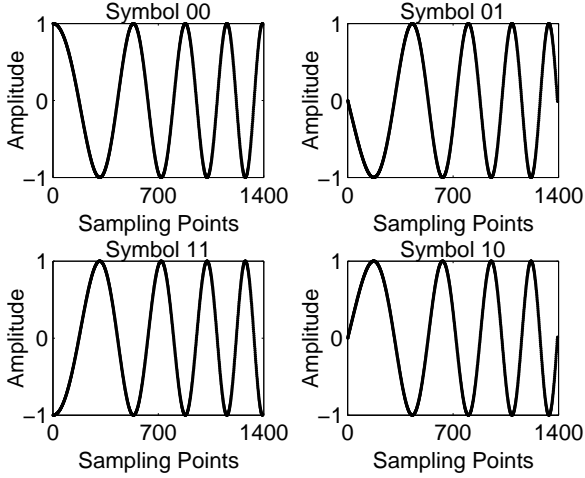}
\caption{QPSK Modulation Based on LFM Carrier-waves.}
\label{fig_sim}
\end{figure}
\subsection{PTRM Equalization}
PTRM is widely used in the underwater acoustic communication
\cite{Stojanovic2008,Shahabudeen2008,Stojanovic2009}, which has
advantage of matching the underwater acoustic channel automatically
without any prior knowledge. PTRM is computational efficient, which
contributes to implement a real-time communication. The schematics
of PTRM is showed in Fig. 5.
\begin{figure}[!t]
\centering
\includegraphics[width=3.4in]{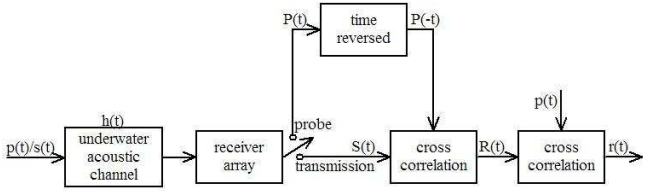}
\caption{PTRM Schematic Diagram.} \label{fig_sim}
\end{figure}

The transmission begins with sending a probe by the transmitter,
denoted as $p(t)$. The probe is an LFM signal. After the multipathed
arrivals finishing, the transmitter sends the data stream, denoted
as $s(t)$. The received probe, denoting as $P(t)$, is transformed
into the time reversed signal, denoted as $P(-t)$. Then the $P(-t)$
is cross-correlated with the received data stream, which denoted as
$S(t)$. The result of the cross-correlation is denoted as $R(t)$.
Finally $R(t)$ is cross-correlated with $p(t)$ to get the
equalization result, denoted as $r(t)$. $r(t)$ is approximative to
$s(t)$. From the foregoing statement, we have
\begin{equation}
\label{eqn_example} P(t) = p(t) \otimes h(t)
\end{equation}
\begin{equation}
\label{eqn_example} S(t) = s(t) \otimes h(t)
\end{equation}
where $h(t)$ is the impulse response of the underwater acoustic
channel, $\otimes$ means cross-correlation.
\begin{equation}
\label{eqn_example} \begin{array}{l}
 R(t) = S(t) \otimes P( - t) = s(t) \otimes h(t) \otimes p( - t) \otimes h( - t) \\
 {\rm{       }}= s(t) \otimes p( - t) \otimes \delta (t)  = s(t) \otimes p( - t), \\
 \end{array}
\end{equation}
\begin{equation}
\label{eqn_example} r(t) = R(t) \otimes p(t) = s(t) \otimes p( - t)
\otimes p(t) = s(t)\otimes \delta (t) = s(t).
\end{equation}

To this end, we find $r(t)$ is $s(t)$, due to the $\delta (t)$
functions we got after PTRM equalization. Overlay of all the signals
got by multi-hydrophone enhances the peak and suppresses the side
lobes, as a result contribute to a better approximative $\delta (t)$
function \cite{Rouseff2001}.

The frame structure of the transmitted signals is showed in Fig. 6.
${T_p}$ is the duration of the probe, ${T_1}$ is the duration of the
guard delay after the probe, ${T_s}$ is the duration of the data
stream, ${T_2}$ is the duration of the guard delay after the data
stream, which is equal to ${T_1}$. $T$ is a symbol period. As the
channel property changes, it is necessary to insert a new probe
after a period of transmission. In order to waiting for the channel
to clear multipath arrivals, a guard delay is inserted between the
data stream and the probe as well as between the probe and the data
stream \cite{Rouseff2001,Yin2008}.
\begin{figure}[!t]
\centering
\includegraphics[width=3.3in]{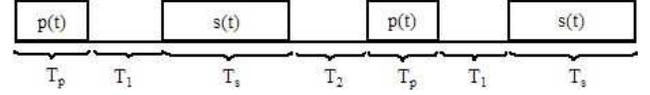}
\caption{Frame Structure of Transmitted Signals.} \label{fig_sim}
\end{figure}
\subsection{Matched Filter Demodulation}
Matched filter is used to demodulate the signals after equalization.
The demodulator at the receiver end stores the time reversed
carrier-waves of all symbols as the reference. The signals after
equalization are cross-correlated with every reference, one by one.
The received signal will be demodulated as the symbol represented by
the reference, which get the maximum peak in the cross-correlation,
here, the maximum peak means the highest similarity. This method is
different from the commonly used method that demodulates the
received carrier-wave as the reference whose cross-correlation is
greater than the decision threshold. Therefore, the larger
difference between different carrier-waves can contribute to a lower
BER. This demodulation method can relax the accuracy of the
equalization and then decrease the complexity of the equalizer.
Furthermore, the method is more appropriate in the complicated
channels, where the received signal can not restore the transmitted
signal very well by the equalization.
\section{CWID modulation}
As the quite severe distortion caused by the underwater acoustic
channel, the received signal after equalization may not be similar
to that of the corresponding reference, but more similar to the
other reference, which leads to a wrong demodulation. The main idea
of the CWID modulation is to increase the difference among different
symbols' carrier-waves. As the difference among different symbols'
carrier-waves increases, the received carrier-wave after
equalization should be more similar to the corresponding reference,
than the others, which contributes to getting a low BER.
\subsection{4-CWID Modulation Based on LFM Carrier-waves}
To show our idea, 4-CWID is used as an illustration. 4-CWID
modulation based on LFM carrier-waves is to divide the LFM signal
carrier-wave into pieces of carrier signals and reorganize the order
of the pieces of carrier signals in order to construct the new
carrier waveform. For illustration, we use a LFM carrier-wave with
the symbol period $T$=0.348ms, the bandwidth 5-18kHz for
illustration. A LFM carrier-wave is showed in Fig. 7.
\begin{figure}[!t]
\centering
\includegraphics[width=2.5in]{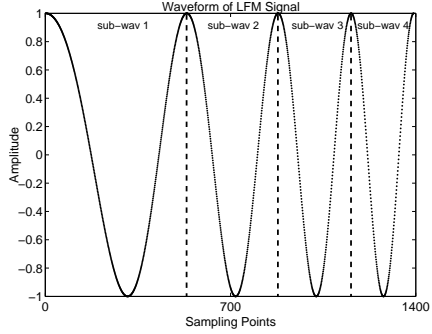}
\caption{LFM Signal with Zero Initial Phase.} \label{fig_sim}
\end{figure}
We split the waveform at every peak of the waveform, to obtain four
pieces of sub-waveforms, marked sequentially from sub-wav 1 to
sub-wav 4, as shown in Fig. 7. We reorganize the sequence of the
sub-waveforms using the rule in Table II in order to get the largest
difference between two different symbols. The main idea of getting
the largest difference between two different symbols is that the
same sub-wav will not appear in the same place of any two different
carrier-waves, for example, sub-wav 1 is located in 1st place in the
carrier-wave of symbol 00, then, sub-wav 1 should not be located in
1st place in the other carrier-waves.
\begin{table}[!t]
\renewcommand{\arraystretch}{1.3}
\caption{Reorganization Sequence of Four Sub-waveforms}
\label{table_example} \centering
\begin{tabular}{c||c||c||c||c}
\hline
\bfseries Symbols & \bfseries 1st place & \bfseries 2nd place & \bfseries 3rd place & \bfseries 4th place\\
\hline\hline
00            & sub-wav 1         & sub-wav 2       & sub-wav 3       & sub-wav 4\\
01            & sub-wav 2         & sub-wav 4       & sub-wav 1       & sub-wav 3\\
10            & sub-wav 3         & sub-wav 1       & sub-wav 4       & sub-wav 2\\
11            & sub-wav 4         & sub-wav 3       & sub-wav 2       & sub-wav 1\\
\hline
\end{tabular}
\end{table}

The 4-CWID carrier waveforms for different symbols are given in Fig.
8. Comparing the waveforms in Fig. 8 with those in Fig. 3 and Fig.
4, we see that the difference of the symbols become larger. The
simulation results in the following part will show this contribute
to the lower BER compared to cosine QPSK and LFM QPSK.
\begin{figure}[!t]
\centering
\includegraphics[width=3.5in]{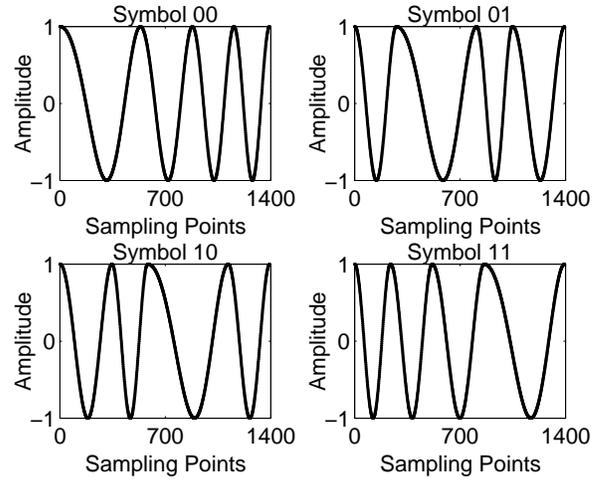}
\caption{4-CWID Modulation Based on LFM Carrier-waves.}
\label{fig_sim}
\end{figure}

Remark 1: we can always find the special point, which make the
sub-wavs connect each other smoothly, to split and reorganize the
waveform in order to obtain the carrier-waves for CWID modulation,
when the initial phase, the symbol period or the bandwidth of the
LFM carrier-waves are changed. This means that the method is a
general method.
\subsection{8-CWID Modulation Based on LFM Carrier-waves}
Select the LFM signals with the initial phase 0 and $\pi$ as the
basic carrier-waves for our 8-CWID modulation. Then carrier-waves
can be obtained by splitting and reorganizing each carrier-wave as
what we did in 4-CWID. The carrier-waves for 8-CWID are given in
Fig. 9. The symbol period of the carrier-waves $T$=0.348ms, the
bandwidth of the carrier-waves is also 5-18kHz.
\begin{figure*}[!t]
\centering
\includegraphics[width=7in]{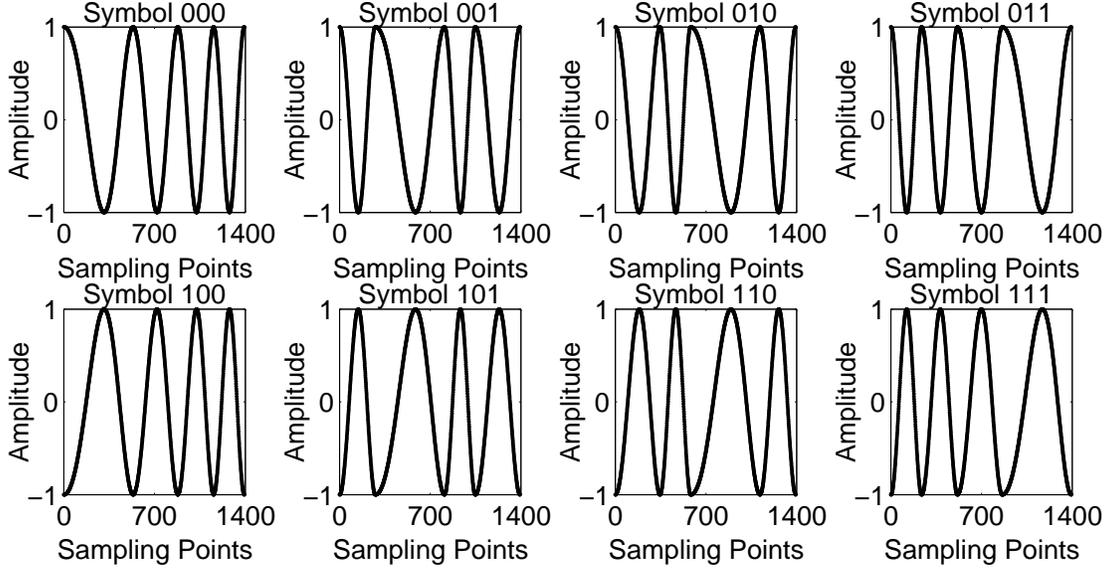}
\caption{8-CWID Modulation Based on LFM Carrier-waves.}
\label{fig_sim}
\end{figure*}
\subsection{16-CWID modulation based on LFM carrier-waves}
Select the LFM signals with the initial phase 0, $\frac{\pi}{2}$,
$\pi$ and $\frac{{3\pi}}{2}$ as the basic carrier-waves for our
16-CWID modulation. Then carrier-waves can be obtained by splitting
and reorganizing the basic carrier-waves as what we did in 4-CWID
and 8-CWID. The carrier-waves for 16-CWID are given in Fig. 10. The
symbol period of the carrier-waves $T$=0.348ms, the bandwidth of the
carrier-waves is 5-18kHz.
\begin{figure*}[!t]
\centering
\includegraphics[width=7in]{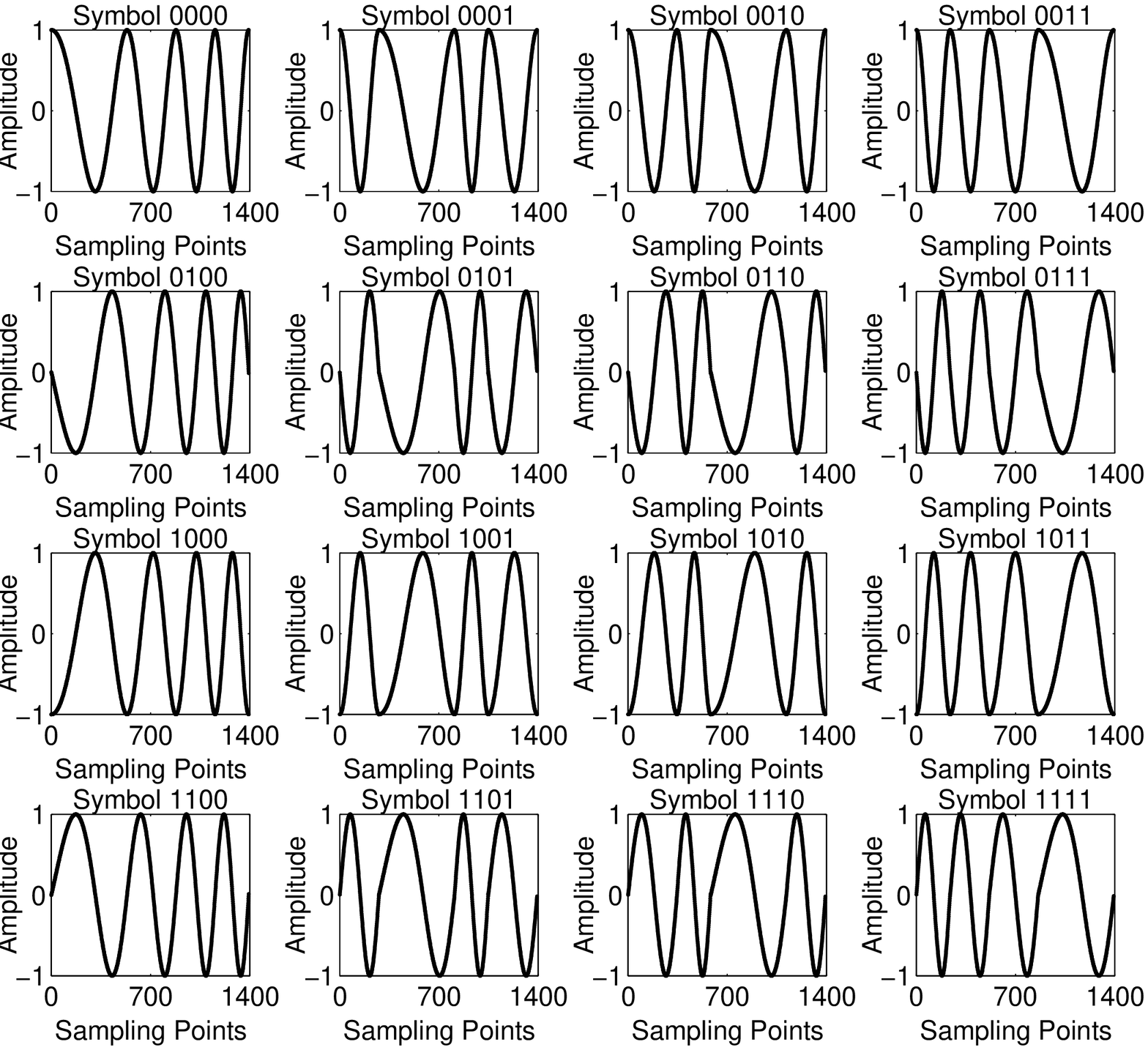}
\caption{16-CWID Modulation Based on LFM Carrier-waves.}
\label{fig_sim}
\end{figure*}

Simulation will be done in the following section to show the result
of the proposed modulation method.
\section{Simulation results}
\subsection{Test of 4-CWID Modulation}
Simulation was done to demonstrate the effectiveness of CWID
modulation. In the simulation, the bandwidth of the LFM
carrier-waves is 5-18kHz. The underwater acoustic channel has calm
surface, smooth bottom, and 30 meters water depth. The sound speed
in the entire propagation region is 1480m/s. The transmission
distance between the transmitter and receiver is 650 meters. The
transmitter hydrophone is deployed at the depth of 15 meters under
water surface. The receiver array, which consists of 9 hydrophones,
is ranged from 8.2 meter to 21.8 meter with a spacing interval of
1.7m. The spatial diversity provided by the array enhances the
performance of PTRM equalization \cite{Rouseff2001}. The transmitter
and receiver are drifting, which causes the Doppler spreading and
shifting. The drifting is decomposed into horizontal drifting and
vertical drifting. The Doppler shifting is set as 10Hz, which equals
to the relative horizontal drifting of 1.3m/s. The statistical
variations of the fading are modeled as independent Rayleigh
processes. The three coefficients of the Rayleigh process for three
eigenrays are ${\sigma_1}=0.5$, ${\sigma_2}=1.0$ and
${\sigma_3}=2.0$. The statistical variations of the time jitter are
modeled as Gaussian processes with zero mean and variance
${\sigma^2}=0.01$, where ${\sigma^2}=0.01$ means the 1-5cm vertical
drifting. The noise is modeled as white noise. The data frame of the
transmitted signals is showed as Fig. 6, where ${T_p}=T=0.348$ms,
${T_1}={T_2}=100$ms since the longest multipath delay spread based
on the first arrival is almost 100ms, ${T_s}=250$ms is to make the
probe and the data stream to be in a coherence time. The ideal
synchronization is assumed in the simulation.

A testing ASCII code stream of "New modulation method for wireless
acoustic communication" are sent in first simulation. The results
showed that the message can be decoded correctly in the receiver.

\subsection{Performance Comparison of Different Modulation Methods}
The simulations of QPSK modulation based on cosine carrier-waves,
LFM QPSK, the method proposed in [8], 4-CWID modulation, 8-CWID
modulation, 16-CWID modulation are performed in this paper. The
underwater acoustic channel properties and the data frame structure
are the same as the case in 5.1. The simulation results are given in
Table III. The SNR of the simulation showed in Table III is 8dB. The
BER is got from transmitting 4000 random bits. The frequency of the
cosine carrier-waves in the simulations is 11.5kHz. The LFM
carrier-waves in the simulations is 5-18kHz. The method in [8] is
designed by autocorrelating a Hamming windowed 5-18kHz LFM signal.
The modulation method used in [8] can be considered as a kind of
DBPSK modulation. The equalizations used in all simulations are
PTRM.
\begin{table*}[!t]
\renewcommand{\arraystretch}{1.3}
\caption{Comparison of Different Methods with SNR equal to 8 dB}
\label{table_example} \centering
\begin{tabular}{c||c||c||c||c||c||c}
\hline
\bfseries             & \bfseries QPSK    & \bfseries LFM QPSK    & \bfseries Method in [8]    & \bfseries 4-CWID     & \bfseries 8-CWID     & \bfseries 16-CWID\\
\hline\hline
Equalization          & PTRM              & PTRM                  & PTRM                       & PTRM                 & PTRM                 & PTRM\\
Transmitting bit rate & 3.19kbit/s        & 3.19kbit/s            & 1.59kbit/s                 & 3.19kbit/s           & 4.78kbit/s           & 6.38kbit/s\\
Bandwidth             & 11.5kHz           & 5-18kHz               & 5-18kHz                    & 5-18kHz              & 5-18kHz              & 5-18kHz\\
BER                   & 0.163             & 0.057                 & 0                          & 0                    & 0                    & 0.092\\
\hline
\end{tabular}
\end{table*}

Figure 11 shows the result of performance comparison of different
modulations under different SNR. The BER of the 4-CWID modulation is
0, when the SNR is larger than -4dB. The BER of the 8-CWID
modulation and the method proposed in [8] are 0, when the SNR is
larger than 4dB. The transmitting bit rate of 4-CWID is the same as
these of the conventional QPSK and the LFM QPSK, whilst, the 4-CWID
possesses much less BER than the conventional QPSK and the LFM QPSK.
When the SNR is larger than -12dB, the proposed 8-CWID has higher
transmitting bit rate and lower BER compared to conventional QPSK,
LFM QPSK and method in [8]. All these show the superiority of the
proposed method.
\begin{figure}[!t]
\centering
\includegraphics[width=3.7in]{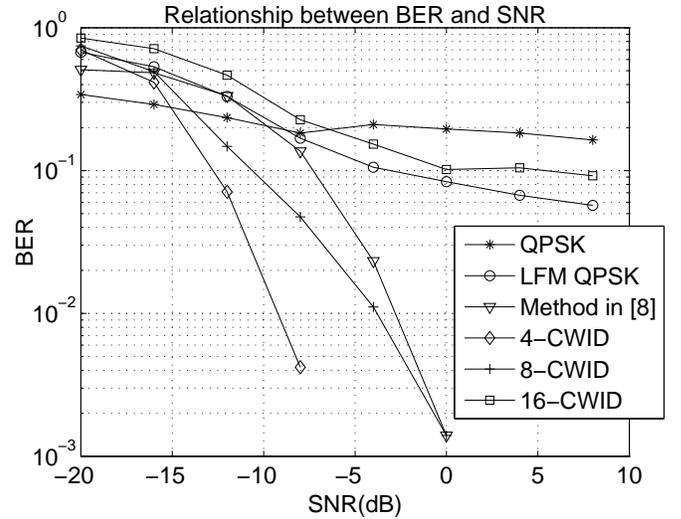}
\caption{Comparison of the Methods in the Paper.} \label{fig_sim}
\end{figure}
\section{Conclusions}
Due to the serious distortion caused by the underwater acoustic
channel, it is very difficult to demodulate the information with
high transmission rate and low BER. To solve this problem, we
propose a novel CWID modulation to increase the difference of the
carrier waveforms of different symbols. The larger difference of the
carrier waveforms contributes to the lower BER. The CWID modulation
is more appropriate for M-ary PSK modulation, which could achieve a
higher transmitting bit rate with a low BER. The bandwidth
efficiency is reduced in CWID modulation, due to the LFM
carrier-waves are used. The reduction of the bandwidth efficiency
brings the higher bit transmission rate and lower BER. The CWID
modulation is more appropriate in complex channels. The bandwidth
efficiency could be improved by optimizing the bandwidth of the LFM
carrier-waves, which will be addressed in the future work.
Comparison of the simulation results shows the superiority of the
proposed methods. CWID modulation can also be used in other media of
communications, for example, free space wireless communication.

\appendices




\ifCLASSOPTIONcaptionsoff
  \newpage
\fi

\bibliographystyle{IEEEtran}
\bibliography{prl_bib_0133}
\end{document}